\documentclass[journal]{IEEEtran}
\usepackage{cite}
\usepackage{amsmath,amssymb,amsfonts}
\usepackage{algorithmic}
\usepackage{graphicx}
\usepackage{textcomp}
\usepackage{subfigure}
\usepackage{caption,setspace}

\ifCLASSINFOpdf
\else
\fi
\begin{document}
%
\title{Broadband Sparse Array Focusing Via Spatial Periodogram Averaging and Correlation Resampling}
%
%
%
%

\author{Yang~Liu,~
        John~R~Buck,~\IEEEmembership{Senior Member,~IEEE}
\thanks{Dr. Yang Liu is with the Consumer Electronics Division, Bose Corporation, 100 the Mountain Rd, Framingham, MA  01701 USA (e-mail: yangliu$\_$acoustics@outlook.com). Dr. John R . Buck is with the Department of Electrical and Computer Engineering, University of Massachusetts Dartmouth, 285 Old Westport Rd, Dartmouth,  MA 02747 USA (e-mail: jbuck@umassd.edu). This material is based upon research supported by the U.S. Office of Naval Research under award numbers N00014-13-1-0230, N00014-17-1-2397 and N00014-18-1-2415.}}

\maketitle
\begin{abstract}
This paper proposes two coherent broadband focusing algorithms for spatial correlation estimation using sparse linear arrays.  Both algorithms decompose the time-domain array data into disjoint frequency bands through discrete Fourier transform or filter banks to obtain broadband frequency-domain snapshots. The periodogram averaging (AP) algorithm starts in the frequency domain by estimating the broadband spatial periodograms for all bands and then averaging them to reinforce the sources' spatial spectral information. Taking inverse spatial Fourier transform of the combined spatial periodogram estimates the focused spatial correlations. Alternatively, the spatial correlation resampling (SCR) algorithm directly computes the spatial correlations for each band and then rescales the spatial sampling rate to align at a focused frequency. The resampled spatial correlations from all frequency bands are then averaged to estimate the focused spatial correlations. The spatial correlations estimated from the AP or SCR algorithms populate the diagonals of a Hermitian Toeplitz augmented covariance matrix (ACM). The focused ACM is the input of a new minimum description length (MDL) based criteria, termed MDL-gap, for source enumeration and the standard narrowband MUSIC algorithm for DOA estimation. Numerical simulations show that both the AP and SCR algorithms improve source enumeration and DOA estimation performances over the incoherent subspace focusing algorithm in snapshot limited scenarios.
\end{abstract}

\begin{IEEEkeywords}
Coherent broadband focusing, Sparse arrays, Augmented covariance matrix, Periodogram averaging, Spatial correlation resampling, Source enumeration, Direction-of-arrival estimation, Snapshots-limited
\end{IEEEkeywords}



%

\section{Introduction}
\label{secIntro}

%
%
%
%
\IEEEPARstart{S}{parse} linear arrays sample a spatial aperture with fewer sensors than required by a standard half-wavelength sampled array.  Many sparse array designs prune or thin a uniform linear array (ULA), so the sparse array sensor locations fall on an underlying half-wavelength lattice \cite{JohnsonBook}. Examples of sparse arrays in this class include minimum redundancy arrays (MRA \cite{MRA}), coprime arrays (CSA \cite{PPCSA}), and nested arrays \cite{PPNested}. These arrays have many array processing applications including source detection \cite{CSADetectionICASSP}, Direction-of-Arrival (DOA) estimation \cite{PPCSAMUSIC}\cite{Aminmultiple} and spatial power spectral density (PSD) estimation \cite{kaushallyaColinear}\cite{YangTSP}. Assuming a large number of snapshots, sparse array processing techniques can localize more sources than sensors by constructing augmented covariance matrices (ACMs) using the second or higher order statistics of the propagating electromagnetic or acoustic field \cite{PPNested}\cite{PillaiORG}. This paper considers the problem of enumerating and estimating the DOAs of more sources than sensors using sparse arrays for temporally broadband signals. 

When the incoming sources are  broadband in temporal frequency, it is possible to combine the spectral information from multiple frequency bands to improve the precision of the spatial correlation estimates. Properly combining data across frequency bands reduces the large number of snapshots required in sparse array processing. This approach will be especially useful in acoustical scenarios, which are often limited in available snapshots due to the relatively slow propagation speed for sound, large array apertures and non-stationary sound fields \cite{BragCox}\cite{Cox}. Assuming the signal obervation time is much longer than the signal correlation times, a commonly used processing approach is to decompose the broadband data into disjoint and uncorrelated narrow frequency bands using Discrete Fourier transform (DFT) or filter banks \cite{VanTrees}. The simplest follow-up step is to estimate the number of sources or their DOAs separately for each frequency band and then average the results across all bands as the final estimate. This method is referred to as \textit{incoherent} signal subspace (ISS) method in th sense that it treats the snapshots from each band as uncorrelated data \cite{ISSM}\cite{MorfISSM}. For source enumeration, the ISS method usually computes the information criteria, such as the Akaike information criterion (AIC \cite{AIC}) or Rissanen's minimum description length criterion (MDL \cite{MDL}), for each band before averaged across all bands to achieve a final estimate \cite{WaxKailathSN}\cite{RajAIC}. For DOA estimation, subspace spectral estimation methods such as MUSIC \cite{MUSIC} are applied on each band and then average the pseudo-spectra across all bands to estimate the source DOAs \cite{ISSM}\cite{HanWidebandSPL}.

While the ISS method works well for broadband signals in high SNR scenarios, the performance can suffer severely for low SNRs and limited snapshots \cite{CSSM}, which frequently occur in underwater acoustical environments. In contrast to the ISS method, the \textit{coherent} signal-subspace (CSS) method exploits the correlations between  signal subspaces at different frequencies and combines the narrowband snapshots to construct a single covariance matrix at a focused frequency \cite{CSSM}\cite{hung1988focussing}. The focused covariance matrix can be estimated with a higher statistical precision reflecting the full time-bandwidth product of the broadband sources \cite{krolik1989multiple}. Narrowband techniques can therefore be applied on the focused covariance matrix with lower thresholds on SNR and snapshots for broadband source enumeration and DOA estimation.


The major challenge in the CSS methods is to design focusing algorithms to align the snapshots across frequency bands to coherently estimate a single covariance matrix. Popular broadband focusing algorithms include rotational signal subspace focusing matrix (RSS, \cite{hung1988focussing}), steered covariance matrices (STCM, \cite{krolik1989multiple}), DFT projection \cite{DFTprojection}, weighted average
of signal subspaces (WAVES \cite{WAVES}), beamforming invariance \cite{BICSSM}  and auto-focusing \cite{autofocusing}. Many of these focusing algorithms require preliminary estimates of the number of sources and their DOAs, which increase the computation cost and bias to the final estimates. Moreover, these algorithms were primarily developed in the context of ULAs and do not apply directly to sparse array data. 

This paper extends two broadband focusing algorithms originally proposed for ULAs to sparse arrays: spatial periodogram averaging \cite{Hinich} and spatial resampling \cite{krolik1990focused}. Neither of these extensions require preliminary DOA estimates for broadband focusing and can be applied on any sparse array geometry with a contiguous coarray region, including MRAs, CSAs and nested arrays. Constructing the ACM using the correlations estimated from the spatial periodogram and spatially resampled correlations offers processing gains for both source enumeration and localization over the ISS approach in \cite{HanWidebandSPL}, especially in low SNR and few snapshots scenarios. 

The rest of this paper is organized as follows. Section \ref{Sec2ULA} discusses the broadband signal model and briefly reviews the ISS method for broadband sparse array processing. Section \ref{Sec3SparseACM} proposes the periodogram averaging (AP) and spatial correlation resampling (SCR) based algorithms for broadband focusing. The focused ACMs from these algorithms are then the inputs for the new MDL-gap source enumeration algorithm and the standard narrowband MUSIC DOA estimator.  The performances of the proposed algorithms are compared with extensive numerical simulations in Section \ref{Sec5Simulations}. Section \ref{conclusion} concludes this paper. 

\section{Incoherent sparse array processing}
\label{Sec2ULA}

This section first describes the array signal model for broadband sources impinging on a sparse linear array and then reviews the incoherent method for broadband sparse array processing. 

\subsection{Wideband signal model}
Assume a sparse linear array with $N$ sensors and $D$ broadband planewave signals impinging on the array from the far field with different DOAs within the visible region  $u_1, u_2,...,u_D \in [-1, 1]$. Here we use the directional cosine $u = \cos(\theta)$ to indicate the source DOA, where $\theta \in [0^o, 180^o]$ is the angle-of-arrival with respect to the array endfire. The signal received by the $n$th sensor at time $t$ can be modeled as 
\begin{equation}
x_n(t) = \sum_{i = 1}^D s_i(t-\tau_n(\theta_i)) + \text{n}_n(t),~n=1,...,N
\end{equation} where $\tau_n(\theta_i)$ is the propagation time delay for the $i$th signal arriving at the $n$th sensor and $\text{n}_n(t)$ is the measurement noise at that sensor. We assume both the signals and noise measured by the sensors are samples of wide-sense stationary and ergodic complex Gaussian processes. The time series at each sensor are divided into $L$ segments. Applying the discrete Fourier transform (DFT) to each segment forms multiple non-overlapping narrow frequency bands, from which we extract the frequency domain phasors at the frequencies of interest $f_1,...,f_M \in [f_{min}, f_{max}]$ \cite{VanTrees}. The segment duration is assumed much longer than the signal correlation time, such that the different DFT bins are statistically uncorrelated. The vector of DFT coefficients (or complex phasors) for all $N$ sensors and the $l$th snapshot at frequency $f_m$ is
\begin{eqnarray}
\label{snapshots}
\textbf{x}_l(f_m) = \textbf{A}(f_m)\textbf{s}_l(f_m) + \textbf{n}_l(f_m),~m &=& 1,...,M \nonumber \\ 
 l &=& 1,...,L, \end{eqnarray}where $\textbf{x}_l(f_m) $ is the $N\times 1$ DFT coefficients vector, $\textbf{A}(f_m)$ is the $N\times D$ array manifold matrix at temporal frequency $f_m$ and $\textbf{s}_l(f_m)$ is the $D\times 1$ source amplitudes vector. The array manifold corresponding to the $n$th element and the $i$th source at frequency $f_m$ is
\begin{equation}
\label{arraymanifold}
[\textbf{A}(f_m)]_{n,i} = e^{j(2\pi f_m d_n/c) u_i},
\end{equation}where $d_n$ is the location of the $n$th element with respect to the array phase center and $c$ is the field propagation speed. The source signal amplitudes are assumed uncorrelated zero-mean and circular complex Gaussians $s_i(f_m) \sim CN(0,\sigma^2_{i,m}), i = 1,...,D$ and uncorrelated from the noise. The additive noise is assumed zero-mean, white, and circular complex Gaussian $\textbf{n} \sim CN(\textbf{0},\sigma_n^2 \textbf{I}_N)$. 


\subsection{Incoherent signal subspace method for sparse arrays}
\label{secISSMULA}
For the broadband signal model in \eqref{snapshots}, the ISS method applies narrowband subspace processing to each frequency band and combines the estimation results across all bands for the final estimate \cite{ISSM}\cite{MorfISSM}. The source enumeration and DOA estimation algorithms are often based on the eigenvalues and eigenvectors of the sample covariance matrices (SCM) computed from each of the complex phasors data in \eqref{snapshots} for a ULA. 

\cite{HanWidebandSPL} extended the ISS method to sparse linear nested arrays. We review here its data processing procedures in the context of finite snapshots, as shown in Fig. \ref{Periodogramblock}(a). For any particular frequency band $f_m$, the narrowband SCM averaged over $L$ snapshots follows 
\begin{equation}
\label{SCM}
\textbf{R}_{xx,m} = \frac{1}{L}\sum_{l=1}^L \textbf{x}_l(f_m) \textbf{x}_l^H(f_m),
\end{equation}where $(\cdot)^H$ denotes Hermitian transpose. Reconstructing the SCM to obtain the $(2P-1)\times 1$ correlation vector corresponding to the contiguous region of the difference coarray
\begin{equation}
\label{SScorr}
\textbf{r}_{m}(k) = \frac{1}{\boldsymbol \eta (k)} \sum_{(n_1,n_2)~\in~\zeta(k)} \left[ \textbf{R}_{xx,m} \right]_{n_1,n_2},  
\end{equation} where $\left[ \textbf{R}\right]_{n_1,n_2}$ selects the $(n_1,n_2)$th element of matrix $\textbf{R}$. mThe set $\zeta(k)$ collects every sensor pair ($n_1,n_2$) separated by the difference coarray index $k = n_1-n_2 \in  [1-P,P-1]$ and $\boldsymbol \eta(k) = |\zeta(k)|$ is the co-array weight equal to the cardinality of the set $\zeta(k)$. Note for different sparse array geometries, the co-array span $P$ will be larger than the number of sensors $N$ by different amounts. To exploit fully the degrees-of-freedom (DOFs) offered by the co-array, apply spatial smoothing (SS) to construct a full-rank and positive semi-definite ACM by \cite{PPNested},
\begin{equation}
\label{SS-ACM}
\textbf{R}_{ss,m} = \frac{1}{P}\sum_{i=1}^P \textbf{v}_m^i (\textbf{v}_m^i)^H,
\end{equation}where $\textbf{v}_m^i$ is a $P\times 1$ vector containing the ($P-k+1$)th through ($2P-k$)th element of $\textbf{r}_{m}(k)$. The spatially smoothed ACM for each frequency band goes through eigenvalue decomposition. The eigenvalues are used to compute the information criteria for each frequency band, which are then averaged across all bands for source enumeration. The ISS method takes the source enumeration  estimate and computes narrowband spatial pseudo-spectra for each frequency band, which are then averaged to obtain a broadband pseudo-spectra used to estimate the DOAs. 

Eq. \eqref{SS-ACM} indicates that SS exploits the fourth-order statistics of the propagating field by averaging the covariance matrices computed from the overlapping subarrays of the co-array correlations. For infinite snapshots, the SS eigenvalues are proportional to the squares of the ensemble eigenvalues for a $P$-element ULA \cite{PPNested}\cite{PPLiuSPL}. Thus, information criteria for source enumeration developed for ULA SCM eigenvalues, which are second moments, are more appropriately applied to the square root of the SS-ACM eigenvalues, and not the eigenvalues themselves as in \cite{HanWidebandSPL}.

For both fully populated and sparse linear arrays, the ISS method works relatively well for broadband sources in high SNR and snapshot rich scenarios \cite{HanWidebandSPL}\cite{CSSM}. However, the source enumeration and localization performance suffers in low SNR scenarios, for sources with gaps in spectral energy such as harmonic sources, and in snapshot limited scenarios. To address these issues, the following section proposes two coherent broadband focusing algorithms for sparse array processing. 


\section{Proposed Coherent Wideband Sparse Array Focusing Algorithms}
\label{Sec3SparseACM}
This section proposes two broadband focusing algorithm for coherent correlation estimations: spatial periodogram averaging (AP) and spatial correlation resampling (SCR). The spatial correlation estimates from either of these two algorithms then populate the diagonals of Hermitian Toeplitz ACMs for subspace processing. The proposed approaches are coherent in the sense that they combine the observed data across all frequency bands to estimate a single broadband ACM from which the number of sources and their DOAs are estimated. In this sense, the frequency averaging occurs with the narrowband spatial correlation functions, which still includes phase terms, in contrast with the incoherent approach which averages only the real-valued information criteria and pseudo-spectra. Both these two algorithms can be applied to any sparse array geometry based on a pruned ULA as long as a contiguous coarray region exists.

\subsection{Periodogram averaging}

\begin{figure*}[!t]
\centering
]{\includegraphics[scale=0.18]{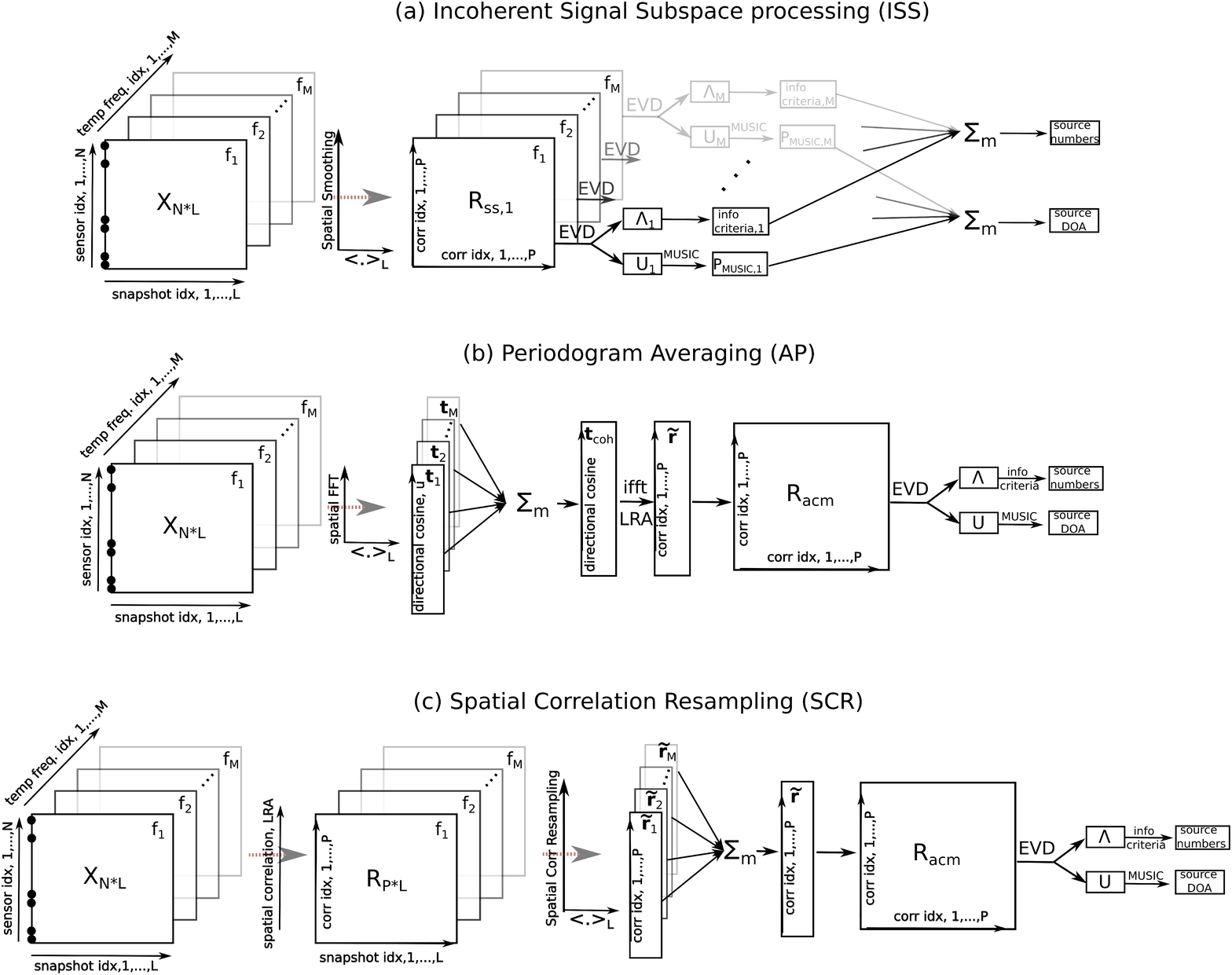}
\label{fig_first_case}}
\caption{Block diagrams for the incoherent signal subspace method (ISS, panel a) and the proposed periodogram averaging (AP, panel b) and spatial correlation resampling (SCR, panel c) algorithms for broadband sparse array source enumeration and DOA estimation. The $N\times L$ matrix $ \bold X = [ \bold x_{1}(f_m),..., \bold x_{N}(f_M)]$ includes the DFT coefficients for all $N$ sensors and $L$ snapshots for frequencies $f_1,...,f_M$}
\label{Periodogramblock}
\end{figure*}
%
%


The spatial periodogram averaging for sparse arrays extends Hinich's broadband beamformer for undersampled ULAs to nonuniform sparse arrays. This approach exploits the frequency diversity obtained through the scanned responses across the signal bandwidth while processing a single ULA \cite{Hinich}. As Fig.~\ref{Periodogramblock}(b) shows, the array frequency snapshot data for each band $f_m,m=1,...,M$ are conventionally beamformed independently via FFT and averaged over all snapshots to estimate the narrowband spatial periodogram $\textbf{t}_m(u)$. The estimated spatial periodogram $\textbf{t}_m(u)$ is the Fourier transform of the estimated spatial auto-correlation function $\textbf{r}_m(k)$ in \eqref{SScorr}, that is routinely used for ACM construciton \cite{PPNested}\cite{PPCSAMUSIC}, weighted by the coarray weights $\boldsymbol \eta (k)$. Specifically, the narrowband periodogram follows
\begin{equation}
\label{narrowbandperiodogram}
\textbf{t}_m(u) = \frac{1}{L} \sum_{l = 1}^L \left|\textbf{w}_m^H(u) \textbf{x}_l(f_m)\right|^2 = \textit{F}_m(\textbf{r}_m(k)\boldsymbol\eta(k)).
\end{equation}where $\textbf{w}_m(u)$ is the conventional beamforming weights vector for frequency $f_m$ at steering direction $u$ (equal to the column vector of the steering matrix $\textbf{A}$ in \eqref{snapshots} for direction $u$) and $\textit{F}_m$ is the spatial Fourier transform operator accounting for the different temporal frequencies $f_m$. In broadband processing, only the true source peaks remain fixed in directional cosine $u$ across different frequency bands, while the grating lobes and sidelobes change their locations in $u$ as the temporal frequency varies. Averaging the periodograms across frequencies constructively reinforces the energy at the true source locations while other sidelobes are relatively attenuated 
\begin{equation}
\label{widebandPeriodogram}
\textbf{t}(u) = \frac{1}{M}\sum_{m=1}^M \textbf{t}_m(u).
\end{equation}
Note that the broadband periodogram in (\eqref{widebandPeriodogram}) has the same functional form as the steered covariance matrix estimate (STCM), which has attractive statisical features expressed in terms of a Wishart characteristic function \cite{krolik1989multiple}. The inverse spatial Fourier transform of the spatial periodogram $\textbf{t}(u)$ estimates the spatial correlation function after normalizing for the coarray weights 
\begin{equation}
\label{correstimatesAP}
\tilde{\textbf{r}}(k) = \frac{\textit{F}^{-1}(\textbf{t}(u))}{\boldsymbol\eta(k)},~k = -(P-1),...,(P-1)
\end{equation}
The estimated broadband correlation function $\tilde{\textbf{r}}(k)$ then populates the diagonals of a Hermitian Toeplitz ACM, as given in Section \ref{subsecACMconstruction}. 

The covariance focusing through periodogram averaging simplifies the coherent broadband processing algorithm while maintaining its advantages in low SNR and
limited snapshot scenarios. Processing broadband data in the beamspace avoids the
complexity of constructing focusing matrices that are commonly required in the
coherent algorithms. Substituting \eqref{narrowbandperiodogram}-\eqref{widebandPeriodogram} into \eqref{correstimatesAP}, the correlation estimates can be written as
\begin{equation}
\tilde{\textbf{r}}(k) = \frac{\textit{F}_c^{-1}\left( \frac{1}{M}\sum_{m=1}^M \textit{F}_m \left(\textbf{r}_m(k)\boldsymbol\eta(k)\right) \right)}{\boldsymbol\eta(k)},
\end{equation} where $\textit{F}_c^{-1}$ is the inverse spatial Fourier transform operator corresponding to the central frequency within the bandwidth. This notation implies that estimating the broadband spatial correlation function through inverse Fourier transform of the averaged spatial periodograms does not account for the temporal frequencies mismatch between frequency bands. To account for this mismatch, it is in general a good practice to perform the inverse Fourier transform at the central frequency of the sources' bandwidth. This is similar to choosing the focusing frequency as the central frequency to reduce DOA estimation bias, as suggested in \cite{KrolikBias}.   

\subsection{Spatial correlation resampling}
Another approach for coherent broadband focusing is through spatial resampling \cite{krolik1990focused}. Spatial resampling exploits the structural characteristic that the array manifold in \eqref{arraymanifold} depends on the source temporal frequencies and the element positions only through their product. By adjusting the spatial sampling intervals of the frequency-domain snapshots as a function of the temporal frequency for each of the frequency bands, it is possible to obtain (nearly) the same array manifold vector at different frequencies. Spatial resampling for broadband processing approaches the performance of the narrowband scenario with a comparable time-bandwidth product \cite{krolik1990focused}.

At first glance, the spatial resampling algorithm previously applied to ULAs cannot be directly applied to sparse array data due to the gaps in the spatial sampling. However, the important insight is that the sparse arrays still provide contiguous and uniformly sampled difference co-array functions. This insight allows us to extend the application of the spatial resampling technique to sparse arrays. Rather than directly resampling the array data, we resample the estimated second-order statistics as a function of spatial lag. To make this insight precise, the spatial correlation between the signals received by sensors located at $d_{n_1}$ and $d_{n_2}$ for a single source with amplitude $s$ from direction $u_i$ can be expressed as
\begin{equation}
E\{x_1(f_m)x_2^*(f_m)\} = s s^* e^{-j(2\pi f_m/c)(d_{n_1}-d_{n_2})u_i}
\end{equation}for frequency band $f_m$. This implies that the array manifold corresponding to the coarray depends on the product of the source temporal frequency $f_m$ and the inter-element spacing $d_{n_1}-d_{n_2}$. Since the contiguous region of the coarray is uniform in spatial lag $k$, applying spatial resampling to the spatial correlations corresponding to this region for all frequency bands will realign the coarray manifolds. The resampling changes the spatial correlation sampling interval for the $m$th band from $d$ to $d_m = d f_0 /f_m$, where $d$ is the physical inter-sensor spacing and $f_0$ is the focus frequency. 

Unlike the periodogram averaging algorithm discussed in Section \ref{subsecAP}, broadband focusing through spatial correlation resampling explicitly accounts for the coarray manifold mismatches between frequency bands due to different temporal frequencies. Fig.~\ref{Periodogramblock}(c) demonstrates the data processing procedures for the SCR algorithm for broadband focusing. Each snapshot data at each frequency band $f_m$ goes through the following procedures:

1) Compute the spatial auto-correlation function by averaging all $L$ snapshots, take the portion corresponding to the contiguous region of the coarray and normalize it by the coarray weights $\boldsymbol \eta(k)$ for unbiased narrowband correlation estimates $\textbf{r}_m(k)$ in \eqref{SScorr}. 

2) Since the correlation estimate is even conjugate symmetric about the coarray center, we apply spatial resampling only to the right half side of the correlation estimate such that $\textbf{z}_m(k) = \textbf{r}_m(k= 0,...,P-1)$ to save computation. 

3) Choose integers $K_m$ and $L_m$ appropriately such that $K_m/L_m = f_m/f_0$, where $K_m$ and $L_m$ are both integers.

4) Upsample $\textbf{z}_m(k)$ by inserting $(K_m-1)$ zeros in between each sample of the correlation estimate such that
\begin{equation}
\textbf{z}'_m(k) = \left\{ \begin{array}{c}
\textbf{z}_m\left(k/K_m\right) ,\text{for}~k = 0, K_m, ..., (P-1)K_m\\ ~~~~~~~~~0~~~~~~,~\text{otherwise.}~~~~~~~~~~~~~~~~~~~~~~~~~
\end{array} \right.
\end{equation}

5) Filter the upsampled correlation function $\textbf{z}'_m(k)$ by a linear phase finite impulse response low pass filter with cut-off frequency of $\min(\pi/K_m, \pi/L_m)$ to obtain the interpolated correlations $\textbf{z}'_{m,\text{intp}}(k)$. Shift or re-index $\textbf{z}'_{m,\text{intp}}(k)$ to obtain the correct set of correlations by accounting for the group delay due to linear phase filtering.

6) Decimate $\textbf{z}'_{m,\text{intp}}(k)$ by a factor of $L_m$ such that $\tilde{\textbf{z}}_m(k) = \textbf{z}'_{m,\text{intp}}(L_m k)$ to obtain the focused spatial correlation function. 

7) Make up for the left half side of the resampled correlation estimates using the even conjugate symmetry property such that 
\begin{equation}
\tilde{\textbf{r}}_m(k) = \left\{ \begin{array}{c}
~~~\tilde{\textbf{z}}^*_m(-k), ~\text{for}~k = -(P-1),...,-1\\
\tilde{\textbf{z}}_m(k), ~\text{for}~k = 0,...,P-1
\end{array} \right.
\end{equation} 

The procedures above are repeated for all snapshots at all frequency bands before averaging across all $M$ frequencies to obtain the coherently combined spatially correlation estimates 
\begin{equation}
\label{SRcorr}
\tilde{\textbf{r}}(k) = \frac{1}{M}\sum_{m=1}^M \tilde{\textbf{r}}_m(k).
\end{equation}  The estimated correlation function $\tilde{\textbf{r}}(k)$ then populates the diagonals of a Hermitian Toeplitz ACM as given in Section \ref{subsecACMconstruction}.

The spatial resampling procedures are essentially the same as time domain resampling, as described in Fig. 4.28 \cite{OppenheimDSPbook}, adapted to spatial correlation functions. It is worth to note that, in theory, the focus frequency can be any value equal to or below the array design frequency to avoid spatial aliasing. However, for practical implementation, we choose to focus at the minimum frequency in band such that $f_0 = f_1$. Resampling in this case corresponds to an interpolation or spatial sampling rate increase by a factor of $K_m/L_m$ at the $m$th frequency band. This makes sure that no extrapolation is needed in Step 5) to guarantee enough correlation samples to decimate in Step 6) in order to maintain the same coarray support for spatial correlation estimates as before resampled.


\subsection{Augmented covariance matrix construction}
\label{subsecACMconstruction}
An alternative approach to SS for ACM construction is through lag redundancy averaging (LRA) \cite{PillaiORG}. This technique exploits the coarray redundancies by averaging all repeated estimates of the spatial correlation function at any given lag from different sensor pairs and then replacing the individual estimates at that lag by their average \cite{noteLRA}\cite{LRAperformance}. As a result, the constructed ACM is populated with correlation estimates with reduced variances. The LRA-ACM is populated with the spatial correlation estimates from either AP \eqref{correstimatesAP} or SCR \ref{SRcorr} following 

\begin{equation}
\label{LRA-ACM}
\textbf{R}_{\text{LRA}} = \left[ \begin{array}{cccc}
\tilde{\textbf{r}}(0) & \tilde{\textbf{r}}(-1) & \cdots & \tilde{\textbf{r}}(1-P) \\ 
\tilde{\textbf{r}}(1) & \tilde{\textbf{r}}(0) & \cdots & \tilde{\textbf{r}}(2-P) \\ 
\vdots & \vdots & \ddots & \vdots \\ 
\tilde{\textbf{r}}(P-1) & \tilde{\textbf{r}}(P-2) & \cdots & \tilde{\textbf{r}}(0)
\end{array} \right].
\end{equation}

The LRA approach constructs a Hermitian Toeplitz ACM from the correlation estimates, although the ACM is positive indefinite. Compared against the SS-ACM, populating the LRA-ACM is more computationally efficient. For the same sparse array data, note that the LRA-ACM exploits the second-order statistics, whereas the SS-ACM exploits the fourth-order statistics of the propagating field. For finite snapshots, the SS-ACM in \eqref{SS-ACM} can be shown explicitly related to the LRA-ACM by \cite{PPLiuSPL}
\begin{equation}
\textbf{R}_{\text{SS}} = \textbf{R}^2_{\text{LRA}}/P.
\end{equation}This implies that $\textbf{R}_{\text{SS}}$ and $\textbf{R}_{\text{LRA}}$ share the same eigen space and the eigenvalues of $\textbf{R}_{\text{SS}}$ are proportional to the square of the eigenvalues of $\textbf{R}_{\text{LRA}}$. For infinite snapshots, the LRA-ACM approaches the ensemble covariance matrix of a fully populated ULA with probability 1 \cite{PillaiACM}. This implies it is more reasonable to use the eigenvalue magnitudes and the eigenvectors of the LRA-ACM rather than the SS-ACM for source enumeration and DOA estimation. 


\subsection{Source Enumeration and DOA estimation}

The ACM constructed in \eqref{LRA-ACM} goes through eigenvalue decomposition, with the eigenvalues sorted in descending order by their magnitudes \cite{PPLiuSPL}
\begin{equation}
|\lambda_1| \geq |\lambda_2| \geq ... \geq |\lambda_k| \geq ... \geq |\lambda_P|,
\end{equation} before computing the information criteria for source enumeration. Rissanen proposed estimating the number of sources as the model order that yields the minimum code length over a range of possible number of sources \cite{MDL}\cite{WaxKailathSN}. The proposed MDL criterion is the sum of the log-likelihood of the maximum likelihood estimator of the model parameters and a bias correction term penalizing over-fitting of the model order
\begin{equation}
\label{MDL}
\text{MDL}(q) = - \log \left( \frac{g_q}{a_q}  \right)^{(P-q)L} + \frac{1}{2}q(2P-q)\log L, 
\end{equation} for the possible number of sources $q = 0, ..., P-1$. The functions $g_q = \prod_{j = q+1}^P |\lambda_j|^{1/(P-q)}$ and $a_q = \frac{1}{P-q}  \sum_{j = q+1}^P |\lambda_j|$ are, respectively, the geometric and arithmetic mean of the $P-q$ smallest eigenvalues of the Wishart distributed SCM. The estimated number of sources is $\hat{q} = \arg \min_q \text{MDL}(q)$. Since the ACM in \eqref{LRA-ACM} does not follow Wishart distribution, there is no theoretical guarantee that the MDL criterion achieves an accurate estimate of the number of sources, especially in under-determined scenarios. 

\cite{LiuBuckMDLgap} modified the standard MDL criterion in \eqref{MDL} and extended its application to the LRA-ACM for enumerating more sources than sensors using narrowband sparse arrays.The new information criterion, termed MDL-gap, is defined as the first-order backward difference of the MDL criterion normalized by the number of snapshots such that

\begin{small}
\begin{eqnarray}
\label{MDlgapcriteria}
\text{MDL-gap}(q) &=& (\text{MDL}(q) - \text{MDL}(q-1))/L \\ \nonumber
&=& -\log \left( \frac{(a_{q-1})^{P-q+1}}{|\lambda_q| (a_{q})^{P-q} } \right) + \frac{P-q+1/2}{L} \log L, 
\end{eqnarray}
\end{small}for the possible number of sources $q = 1,...,P-1$. The detected source number is $\hat{q} = \arg \min_q \text{MDL-gap}(q)$. Since the MDL-gap criterion showed improved performance over MDL in enumerating more sources than sensors in the narrowband scenarios \cite{LiuBuckMDLgap}, we here extend its application to the LRA-ACM in \eqref{LRA-ACM} for broadband sources. 

Assuming the number of sources is accurately estimated, the DOA estimation is performed by directly applying the standard narrowband spectral MUSIC algorithm \cite{MUSIC} to the coherently constructed ACM. Specifically, the  eigenvectors corresponding to the $P-D$ least significant eigenvalues of the ACM are extracted to estimate the noise subspace 
\begin{equation}
\label{coherentnoisesub}
\textbf{V}_\text{coh}^{\perp} = [\textbf{v}_{D+1},\textbf{v}_{D+2},...,\textbf{v}_P].
\end{equation}
Since the source manifold vectors at the focused frequency 
\begin{equation}
\textbf{a}(u_i)  = [1,...,e^{j(2\pi f_0 kd/c)u_i},...,e^{j(2\pi f_0 Pd/c)u_i}]
\end{equation} for each source $i=1,...,D$ are orthogonal to the noise subspace spanned by $\textbf{V}_\text{coh}^{\perp}$, the MUSIC spectra computed as 
\begin{equation}
\label{cohMUSIC}
P_{\text{coh}}(u) = \frac{1}{\textbf{a}(u)^H\textbf{V}_\text{coh}^{\perp} (\textbf{V}_\text{coh}^{\perp})^H \textbf{a}(u)},
\end{equation} will show $D$ peaks at the source locations. The source DOAs are then estimated by searching for the highest $D$ peaks in the coherently estimated MUSIC spectra.

\section{Comparative Simulation Results and Performance Analysis}
\label{Sec5Simulations}
This section compares the performance of the proposed AP and SCR based broadband focusing algorithms for source enumeration and DOA estimation in numerical simulations. These approaches are compared against the ISS processing in scenarios with relatively few snapshots. All simulations in this section model the source amplitudes as uncorrelated, complex Gaussians with equal power occupying a bandwidth of 40 Hz around the central frequency of 100 Hz. The broadband sources are decomposed evenly into 41 narrowband components via FFT within the bandwidth. As a benchmark, we compare all simulations against the narrowband (NB) case with comparable time-bandwidth product to the broadband sources. This means the narrowband sources has 41 times more snapshots than the broadband sources. This comparison with the narrowband case makes clear the performance cost paid by the focusing operations where the proposed broadband algorithms combine information across the frequency band.  

For demonstration purposes, we compare a MRA with 6 sensors at locations $[1,2,5,6,12,14]d$. This array offers a contiguous coarray region spanning $k \in [-13,13]$. The fundamental inter-element spacing of the MRA is $d = \lambda/2$, where $\lambda$ is the spatial wavelength at the central frequency $f =$ 100 Hz. The sensor SNR level is defined as the ratio between the power of each source signal to the noise power at a single sensor. The noise is assumed both temporally and spatially white and complex Gaussian occupying the same bandwidth as the sources, uncorrelated among the sources and also uncorrelated between each pair of sensors. The following simulations consider two scenarios focusing on different perspectives. The first is an over-determined scenario to demonstrate the proposed algorithms' capability to resolve closely spaced sources. The second is an under-determined scenario to demonstrate the proposed algorithms' capability to enumerate and localize more sources than sensors.  

\subsection{Resolving two closely-spaced sources}

\begin{figure}
  \centerline{\includegraphics[width=10cm]{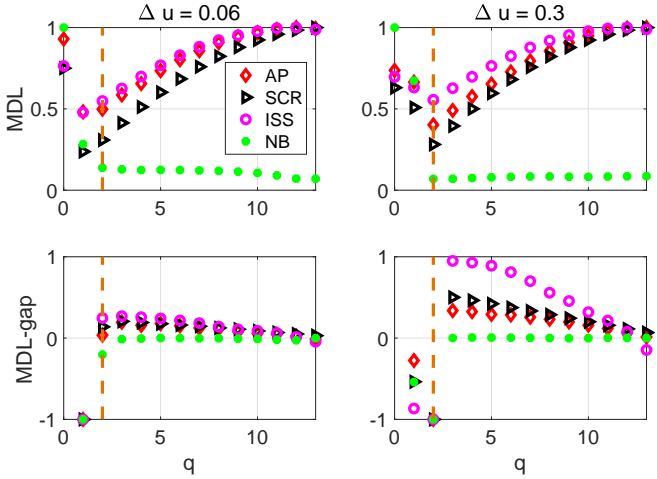}}
\caption{Comparing the sample realizations of MDL and MDL-gap criteria for the AP, SCR, ISS, and equivalent narrowband scenarios for 2 uncorrelated sources. The 2-sources are separated by $\Delta u = 0.06$ on the left column and $\Delta u = 0.3$ on the right column. All simulations assume equal power sources with sensor level SNR = 0 dB using 3 snapshots/sensor for broadband sources and 123  snapshots/sensor for narrowband sources. The results imply all algorithms struggle to enumerate closely spaced sources but start to enumerate correctly when the sources are further separated. }
\label{SampleRealization2sources}
\end{figure}

In the two-source scenario, we first evaluate the performance of the 4 approaches for
source enumeration using the MDL and MDL-gap criteria. Fig. \ref{SampleRealization2sources} compares
the sample realizations of the information criteria as a function of possible number
of sources. All information criteria are normalized by their maximum magnitudes
respectively for demonstration purpose. All simulations use 3 snapshots/sensor for
the broadband approaches and equivalently, 123 snapshots/sensor for the narrowband
sources. There are two sources D = 2 arriving from directions u = [0, 0.06] for the left column and u = [0, 0.3] for the right column. The true number of sources $D=2$ is indicated by orange vertical dashed lines in all panels. For simplicity, all sources
are assumed equal power with sensor level SNR = 0 dB. For the two-source case, all information criteria show minima at D = 2, which implies that all algorithms
are able to estimate the true number of sources. The sample realization results indicate all algorithms struggle to enumerate closely spaced sources but start to enumerate correctly when the sources are further separated.

\begin{figure}
  \centerline{\includegraphics[width=9cm]{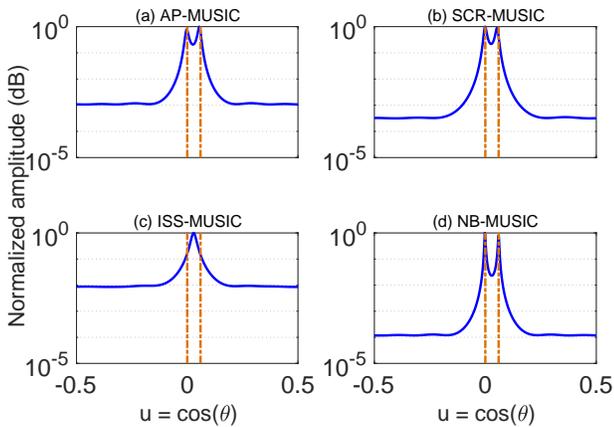}}
\caption{Comparing the (a) AP,  (b) SCR, (c) ISS, and (d) equivalent narrowband MUSIC pseudo-spectra for two uncorrelated sources with DOAs $u = [0,0.06]$ indicated by vertical dashed lines.  All simulations assume equal power sources with sensor level SNR = 0 dB and 3 snapshots per sensor for each of the 41 frequency bands. The equivalent narrowband case uses 123 snapshots/sensor. The MUSIC spectra imply the proposed AP and SCR approaches are more capable of resolving closely spaced sources than the ISS approach. }
\label{MUSICMRA}
\end{figure}

Assuming the number sources are accurately estimated, Fig. \ref{MUSICMRA}  compares the MUSIC pseudo-spectra of AP, SCR, ISS and the equivalent NB scenario in panels ($a$-$d$) for two closely spaced uncorrelated sources with DOAs at $u = [0,0.06]$, which are within the Rayleigh resolution limit $\Delta u = 0.13$ calculated based on the MRA co-array aperture. The two sources are assumed equal power with sensor level SNR = 0 dB and 3 snapshots/sensor for each of the 41 narrow bands. The equivalent NB case uses 123 snapshots/sensor. Note that the AP, SCR and NB MUSIC spectra all show two discernible peaks near the true DOAs indicated by vertical orange dashed lines. However, the ISS approach fails to resolve these two sources, showing only one unique peak in between the true DOAs instead. The MUSIC spectra imply the proposed AP and SCR approaches are more capable of resolving closely spaced sources than the ISS approach.

\begin{figure}
  \centerline{\includegraphics[width=9cm]{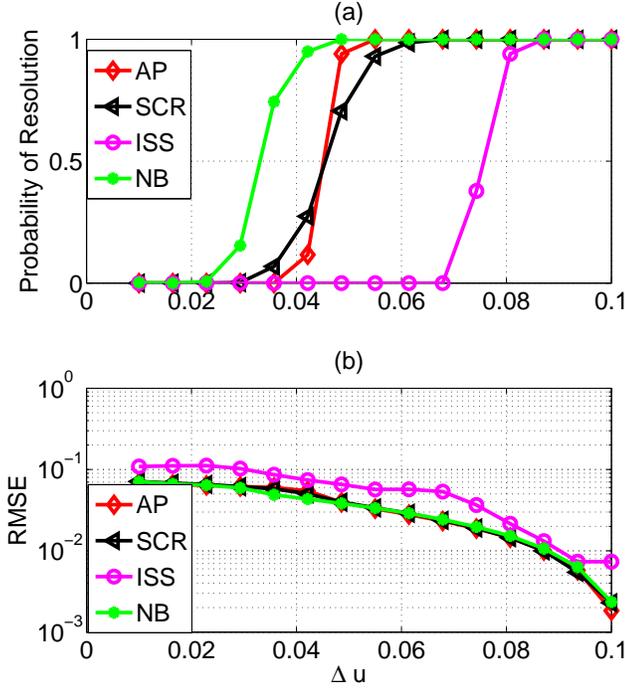}}
\caption{Comparing (a) the probability of resolution and (b) the RMSE of DOA estimates as a function of the spacing between 2 uncorrelated equal power sources with SNR = 0 dB. One source is fixed at broadside and the other source is located away from broadside by $\Delta u$ between [0.01, 0.1]. The simulations for broadband sources use 5 snapshots/sensor and the equivalent narrowband sources use 205 snapshots per sensor. The results indicate the proposed AP and SCR approaches are capable of resolving more closely spaced sources and achieving higher DOA estimate precision than the ISS approach.}
\label{ProbResoRMSEAgaisntDeltaU2sources}
\end{figure}

To characterize rigorously the resolvability and DOA estimate errors of the two closely spaced sources, we compare the 4 approaches on their probabilities of resolution \cite{kaveh1986statistical} and average root mean square errors (RMSE) for all estimated DOAs against the source separation $\Delta u$, source SNR and number of snapshots/sensor. The DOA estimate performance is characterized by the 
\begin{equation}
\text{RMSE} = \sqrt{\sum_{d=1}^D \sum_{j=1}^{J} (\hat{u}_d(j) - u_d)^2/DJ },
\end{equation} where $\hat{u}_d(j)$ is the estimated DOA using the MUSIC algorithm for the $d$-th source in the $j$-th Monte Carlo trial with $d = 1,...,D$ and $j = 1,...,J$. All simulations results are averaged over $J = 500$ independent Monte Carlo trials. Fig. \ref{ProbResoRMSEAgaisntDeltaU2sources}(a) compares the probability of resolution of these 4 approaches as a function of the spacing $\Delta u$ between two sources for SNR = 0 dB and 5 snapshots/sensor. AP and SCR have close performance in resolving two closely spaced sources, which are both worse than the NB case. However, the AP and SCR approaches outperform the ISS approach in their ability to resolve more closely spaced sources. Fig. \ref{ProbResoRMSEAgaisntDeltaU2sources}(b) compares the RMSE for the DOA estimates. For all approaches, the RMSEs decrease as the separation between the two sources increases. The AP, SCR and NB have similar RMSEs, which are lower than the RMSE using the ISS approach.

\subsection{Enumerating/Localizing more sources than sensors}


One major advantage that sparse arrays offer over the fully populated arrays is the capability of localizing more sources than sensors \cite{PillaiORG}. This section explores the advantages of the proposed AP and SCR approaches in enumerating and estimating more broadband sources than sensors over the ISS approach. We again use the same 6-element MRA as in the previous section, but with 9 uncorrelated equal power sources: 1 at broadside, 4 uniformly spaced in $\theta = (90^o, 135^o]$ and the other 4 uniformly spaced in $u = (0,0.7]$. 

\begin{figure}
  \centerline{\includegraphics[width=10cm]{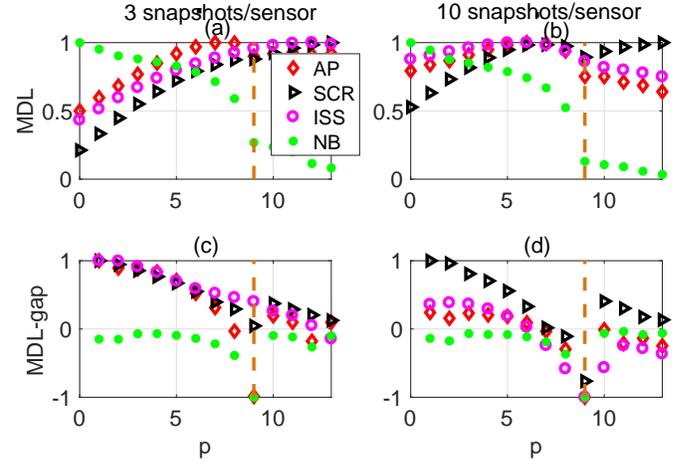}}
\caption{Comparing the sample realizations of MDL and MDL-gap criteria for the AP, SCR, ISS, and equivalent narrowband scenarios for 9 uncorrelated sources. All simulations assume equal power sources with sensor level SNR = 0 dB and 3 snapshots per sensor on the left column and 10 snapshots per sensor on the right column. The results imply that MDL struggles to enumerate more sources than sensors regardless of the number of snapshots available. However, using the MDL-gap criteria, the proposed AP and SCR approaches require fewer snapshots than ISS for correct source enumeration.}
\label{SampleRealization9sources}
\end{figure}

We first evaluate the performance of the 4 approaches for source enumeration using the MDL and MDL-gap criteria. Fig. ~\ref{SampleRealization9sources} compares the sample realizations of the criteria as a function of possible number of sources. All information criteria are normalized by their maximum magnitudes respectively for demonstration purpose. The simulations in the left column of panels $(a,c)$ use 3 snapshots/sensor for the broadband source and equivalently, 123 snapshots/sensor for the narrowband source. The simulations in the right column of panels $(b,d)$
use 10 snapshots/sensor for the broadband source and equivalently, 410 snapshots/sensor for the narrowband source. For all panels, the true number of sources D = 9 is indicated by vertical orange dashed lines. For simplicity, all sources are assumed equal power with sensor level SNR = 0 dB. Panel (a) shows that when the number of sources D = 9 exceeds the number of sensors N = 6, none of the approaches exhibits a minimal MDL value at $\hat{D}$ = 9. Panels (c) shows that the AP, SCR and NB approaches show minimal MDL-gap values at $\hat{D}$ = 9. However, the ISS approach is not able to estimate $\hat{D}$ = 9 using either criteria at the modest snapshots level. When the number of snapshots increases, panel (b) shows the MDL still fails to estimate $\hat{D}$ = 9 for all four methods of constructing the ACM. However, panel (d) shows that all approaches using the MDL-gap criterion are able to correctly estimate the source number $\hat{D}$ = 9. These simulations imply that the AP and SCR approaches are capable of enumerating more sources than sensors in relatively few snapshots using MDL-gap. However, at least in this example, the ISS approach requires relatively large number of snapshots to achieve an accurate enumeration of more sources than sensors using MDL-gap. 

\begin{figure}
\begin{center}
\includegraphics[width=9.2cm]{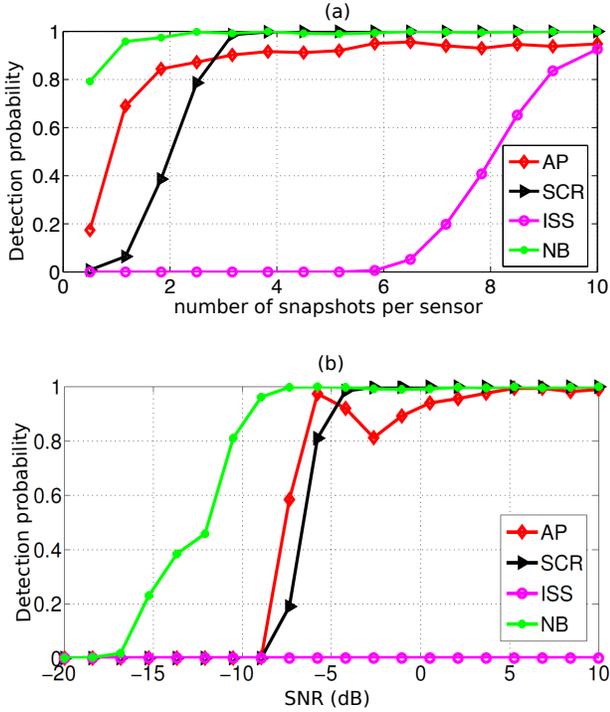} 
\caption{Comparing the probability of correctly enumerating the number of sources using the MDL-gap criterion for different approaches (a) as a function of the number of snapshots per sensor for fixed sensor level SNR = 0 dB and (b) as a function of sensor level SNR for a fixed 5 snapshots per sensor. There are 9 equal power sources impinging on the 6-element MRA. The results indicate the AP and SCR approaches require fewer snapshots and lower SNR than the ISS approach for source enumeration. }
\label{ProbDetection}
\end{center}
\end{figure}

To quantify rigorously the performance of the proposed AP and SCR approaches in enumerating more sources than sensors, Fig.  \ref{ProbDetection}(a) compares the probability of correct enumeration using MDL-gap against snapshots/sensor and Fig. \ref{ProbDetection}(b) against sensor level SNR. The detection probability is calculated as the number of Monte Carlo trials correctly estimating $\hat{D}= 9$ sources, normalized over a total of 500 trials. The sensor level SNRs are the same of 0 dB for all 9 sources for the simulations in panel (a). The simulation results show that the detection probabilities using all approaches increase as the numbers of snapshots increase. In particular, the AP and SCR approaches require much lower numbers of snapshots than the ISS approach to achieve a high detection probability. AP has higher detection probability than SCR for less than 2 snapshots/sensor, but doesn't converge to 1 as fast as SCR. In contrast, ISS requires 6 snapshot/sensor to start detecting all sources and 10 snapshots/sensor to achieve a detection probability above $90\%$. Panel (b) evaluates the detection probability as a function of sensor level SNR. The number of snapshots/sensor is fixed as 5 for the broadband source and 205 for the equivalent NB source. The simulation results show that the NB approach requires the lowest SNR level to start correctly detecting all sources. The AP and SCR approaches require SNR of -9 dB to start detecting all sources. ISS is not  able to enumerate all sources for all SNRs with only 5 snapshots per sensor available.

\begin{figure}
  \centerline{\includegraphics[width=9.8cm]{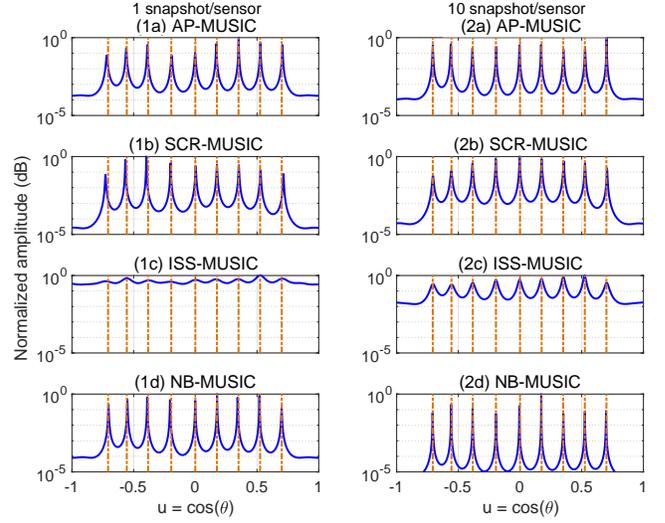}}
\caption{Comparing the (a) AP, (b) SCR, (c) ISS, and (d) equivalent narrowband MUSIC pseudo-spectra for 9 broadband sources with 1 snapshots/sensor (left column) and 10 snapshots/sensor (right column) for the broadband sources. The AP and SCR MUSIC spectra show sharper peaks than the ISS MUSIC spectra for the same number of snapshots, indicating more precise DOA estimation. }
\label{MUSICMRA9sources}
\end{figure}

\begin{figure}
  \centerline{\includegraphics[width=9.5cm]{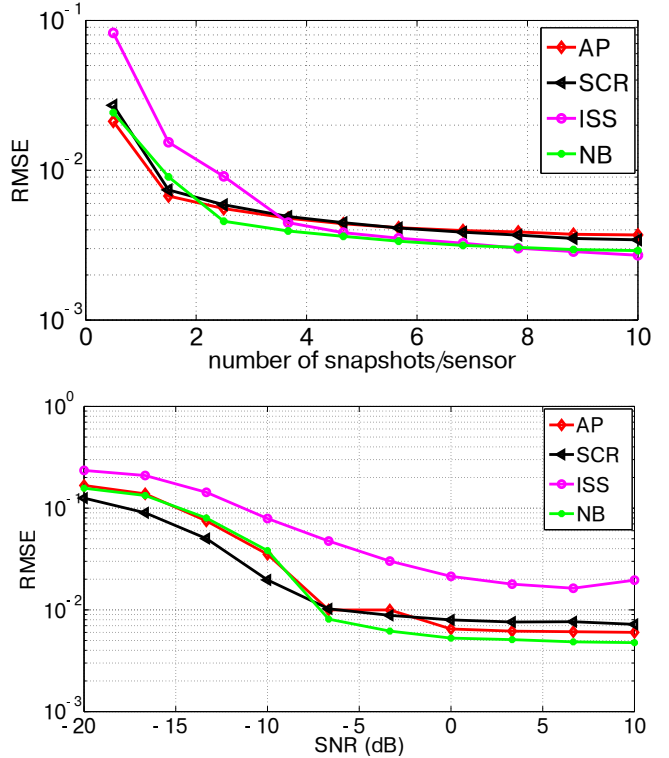}}
\caption{Comparing the RMSE of DOA estimates against (a) snapshots/sensor with fixed SNR = -5 dB (a) and against (b) SNR with fixed 1 snapshots/sensor for 9 uncorrelated equal power sources. The results indicate the AP and SCR algorithms achieve lower RMSE than the ISS algorithm in low snapshots and SNR scenarios.}
\label{MRARMSE9sources}
\end{figure}

Assuming the number of sources is correctly estimated, we explore the DOA estimation performances of the AP and SCR approaches for scenarios with more sources than sensors. Fig. \ref{MUSICMRA9sources} $(1a-1d)$ compare the MUSIC pseudo-spectra of AP, SCR, ISS and the equivalent NB approaches for 9 sources with DOAs indicated by vertical orange dashed lines. All sources are assumed equal power with sensor level SNR = 0 dB and 1 snapshots/sensor for each of the 41 frequency bands. The equivalent narrowband case uses 41 snapshots/sensor.  Note that the AP, SCR and NB MUSIC spectra all show discernible peaks near the true DOAs. However, the ISS approach shows very shallow (smeared) peaks in its MUSIC spectra and misses detecting some sources. Panels $(2a-2d)$ compare the MUSIC pseudo-spectra of AP, SCR, ISS algorithms for 10 snapshots/sensor for the broadband sources and equivalently, 410 snapshots/sensor for the narrowband scenario. When the number of snapshots increases, the MUSIC spectra for all algorithms have sharper peaks at the true DOAs. However, the MUSIC spectra for ISS algorithm is still shallower than the other 3 algorithms. 

Fig. \ref{MRARMSE9sources}(a) compares the RMSEs of all approaches averaged over 500 Monte Carlo trials against number of snapshots/sensor for the 9 sources with fixed SNR = -5 dB. All RMSEs decrease as the number of snapshots increases. AP and SCR have almost identical RMSEs, which are lower than ISS for less than 3 snapshots/sensor, and slightly higher than ISS for above 4 snapshots/sensor. ISS converges to the narrowband scenario closer than both AP and SCR for above 4 snapshots/sensor. Fig.  \ref{MRARMSE9sources}(b) compares the RMSEs of all approaches against sensor level SNR for 9 sources with 1 snapshot/sensor for the broadband scenario, and equivalently 41 snapshots/sensor for the NB scenario. All RMSEs decrease as SNR level increases. AP and NB have very close RMSEs for the SNR range considered, which are both close to the SCR. The ISS has strictly greater RMSE than AP and SCR for all SNR levels due to the low number of snapshots available. These simulations imply that the AP and SCR approaches have advantages over the ISS approach in enumerating and estimating the DOAs of more broadband sources than sensors, especially in low SNR and relatively few snapshots scenarios.

\section{Conclusion}
\label{conclusion}
This paper proposed new coherent broadband focusing algorithms for sparse linear array processing. The proposed algorithms extend the concepts of periodogram averaging and spatial resampling developed for ULAs to the correlation estimates for any sparse array geometries. By averaging the spatial periodograms across multiple narrow frequency bands, the sources' spectral information are constructively reinforced in the beamspace. Alternatively, spatial resampling of the correlation estimates from different frequency bands realigns the co-array manifold mismatches due to distinct temporal frequencies between frequency bands. The improved statistical precision in the correlation estimates offered by both broadband periodogram averaging and spatial correlation resampling construct augmented covariance matrices with higher statistical precision than processing only one frequency band with the same number of snapshots. The broadband algorithms usually pay small penalty for coherent focusing when compared with a narrowband algorithm with the same total number of measurements. The new algorithms proposed in this paper addressed the challenges of sparse array processing in low SNR and snapshot-limited environments. Treating data from other frequency bands as additional snapshots inherently reduces the number of snapshots usually required compared to the narrowband scenarios to achieve a given estimate precision. Improving the precision of spatial correlation estimate benefits the sparse array source enumeration and localization tasks in underwater sonar systems with practically challenging scenarios due to slow speed of sound propagation, relatively large aperture and non-stationary fields.


%

%
%
%
%
  \section*{Acknowledgment}

This material is based upon research supported by the U.S. Office of Naval Research under award numbers N00014-13-1-0230, N00014-17-1-2397 and N00014-18-1-2415.

\ifCLASSOPTIONcaptionsoff
  \newpage
\fi


\begin{thebibliography}{1}

\bibitem{JohnsonBook}
D.H. Johnson and D.E. Dudgeon, \textit{Array signal processing: concepts and technique.} Simon \& Schuster, New Jersey, 1992.

\bibitem{MRA}
A. Moffet, ``Minimum-redundancy linear arrays,'' \textit{IEEE Trans. Antennas Propag.}, vol. AP-16, no. 2, pp. 172-175, 1968.

\bibitem{PPCSA} 
P.P. Vaidyanathan and P. Pal, ``Sparse sensing with co-prime samplers and arrays,'' \textit{IEEE Trans. Signal Process.}, vol. 59, no. 2, pp. 573-586, 2011.

\bibitem{PPNested}
P. Pal and P.P. Vaidyanathan, ``Nested arrays: a novel approach to array processing with enhanced degrees of freedom,'' \textit{IEEE Trans. Signal Process.}, vol. 58, no. 8, pp. 4167-4181, 2010. 

\bibitem{CSADetectionICASSP} 
K. Adhikari and J.R. Buck, ``Gaussian signal detection by coprime sensor arrays,'' \textit{2015 IEEE International Conference on Acoustics, Speech and Signal Processing (ICASSP)}, April 2015.

\bibitem{PPCSAMUSIC}
P. Pal and P.P. Vaidyanathan, ``Coprime sampling and the MUSIC algorithm,'' in \textit{Proc. IEEE Digital Signal Processing Workshop and IEEE Signal Processing Education Workshop}, pp. 289-294, 2011.

\bibitem{Aminmultiple}
E. BouDaher, Y. Jia, F. Ahmad and M.G. Amin, ``Multi-frequency co-prime arrays for high-resolution direction-of-arrival estimation,'' \textit{IEEE Trans. Signal Process.}, vol. 63, no. 14, pp. 3797-3808, 2015.

\bibitem{kaushallyaColinear}
K. Adhikari and J.R. Buck, ``Spatial spectral estimation with product processing of a pair of colinear arrays,'' \textit{IEEE Trans. Signal Process.}, vol. 65, no. 9, pp. 2389-2401, 2017. 

\bibitem{YangTSP}
Y. Liu and J.R. Buck, ``Gaussian source detection and spatial spectral estimation using a coprime sensor array with the min processor,'' \textit{IEEE Trans. Signal Process.}, vol. 66, no. 1, pp. 186-199, 2018. 

\bibitem{PillaiORG}
S.U. Pillai, Y. Bar-Ness and F. Haber, ``A new approach to array geometry for improved spatial spectrum estimation,'' \textit{Proc. IEEE}, vol. 73, no. 10, pp. 1522-1524, 1985.

\bibitem{BragCox}
A.B. Baggeroer and H. Cox, ``Passive sonar limits upon nulling multiple moving ships with large aperture arrays,'' in \textit{Proc. 33rd Asilomar Conference on Signals, Systems and Computers}, vol. 1, pp. 103-108, 1999.

\bibitem{Cox}
H. Cox, ``Adaptive beamforming in non-stationary environments,'' in \textit{Proc. 36th Asilomar Conference on Signals, Systems and Computers}, vol. 1, pp. 431-438, 2002.

\bibitem{VanTrees}
H.L. V. Trees, \textit{Detection, Estimation and Modulation Theory Part IV: Optimum Array Processing.} New York: Wiley, 2002.

\bibitem{ISSM}
M. Wax, T-J. Shan and T. Kailath, ``Spatio-temporal spectral analysis by eigenstructure methods,'' \textit{IEEE Trans. Acoust. Speech Signal Process.}, vol. 32, no. 4, pp. 817-827, 1984.

\bibitem{MorfISSM}
G. Su and M. Morf, ``The signal subspace approach for multiple wide-band emitter location,'' \textit{IEEE Trans. Acoust. Speech Signal Process.}, vol. 31, no. 6, pp. 1502-1522, 1983.

\bibitem{AIC}
H. Akaike, ``A new look at the statistical model identification,'' \textit{IEEE Trans. Autom. Control}, vol. 19. no. 6, pp. 716-723, 1974.

\bibitem{MDL}
J. Rissanen, ``Modeling by shortest data description,'' \textit{Automatica}, vol. 14, pp. 465-471, 1978.

\bibitem{WaxKailathSN}
M. Wax and T. Kailath, ``Detection of signals by information theoretic criteria,'' \textit{IEEE Trans. Acoust. Speech Signal Process.}, vol. 33, no. 2, pp. 387-392, 1985.

\bibitem{RajAIC}
R.R. Nadakuditi and A. Edelman, ``Sample eigenvalue based detection of high-dimensional signals in white noise using relatively few samples,'' \textit{IEEE Trans. Signal Process.}, vol. 56, no. 7, pp. 2625-2638, 2008.

\bibitem{MUSIC}
R.O. Schmidt, ``Multiple emitter location and signal parameter estimation,'' \textit{IEEE Trans. Antennas Propag.}, vol. 34, no. 3, pp. 276-180, 1986.

\bibitem{HanWidebandSPL}
K. Han and A. Nehorai, ``Wideband Gaussian source processing using a linear nested array,'' \textit{IEEE Signal Process. Lett.}, vol. 20, no. 11, pp. 1110-1113, 2013.


\bibitem{CSSM}
H. Wang and M. Kaveh, ``Coherent signal-subspace processing for the detection and estimation of angles of arrival of multiple wide-band sources,'' \textit{IEEE Trans. Acoust. Speech Signal Process.}, vol. 33, no. 4, pp. 823-831, 1985.

\bibitem{hung1988focussing}
H. Hung and M. Kaveh, ``Focusing matrices for coherent signal-subspace processing,'' in \textit{IEEE Trans. Acoust. Speech Signal Process.}, vol. 36, no. 8, pp. 1272-1281, 1988. 

\bibitem{krolik1989multiple}
J. Krolik and D. Swingler, ``Multiple broad-band source location using steered covariance matrices,'' in \textit{IEEE Trans. Acoust. Speech Signal Process.}, vol. 37, no. 10, pp. 1481-1494, 1989. 

\bibitem{DFTprojection}
M. Allam and A. Moghaddamjoo, ``Two-dimensional DFT projection for wideband direction-of-arrival estimation,'' in \textit{IEEE Trans. Signal Process.} vol. 43, no. 7, pp. 1728-1732, 1995.

\bibitem{WAVES}
E. Di Claudio and R. Parisi, ``WAVES: weighted average of signal
subspaces for robust wideband direction finding,'' \textit{IEEE Trans.
Signal Process.}, vol. 49, no.10, pp. 2179-2191, 2001.

\bibitem{BICSSM}
T-S. Lee, ``Efficient wideband source localization using beamforming invariance technique,'' \textit{IEEE Trans. Signal Process.}, vol. 42, no. 6, pp. 1376-1387, 1994.


\bibitem{autofocusing}
P. Pal and PP. Vaidyanathan, ``A novel autofocusing approach for estimating directions-of-arrival of wideband signals,'' in \textit{43rd Asilomar Conference on Signals, Systems and Computers}, pp. 1663-1667, 2009.

\bibitem{Hinich}
M.J. Hinich, ``Processing spatially aliased arrays,'' \textit{J. Acoust. Soc. Am.}, vol. 64, no. 3, pp. 792-794, 1978. 

\bibitem{krolik1990focused}
J. Krolik and D. Swingler, ``Focused wide-band array processing by spatial resampling,'' in \textit{IEEE Trans. Acoust. Speech Signal Process.}, vol. 38, no. 2, pp. 356-360, 1990.

\bibitem{PPLiuSPL}
C.-L. Liu and P.P. Vaidyanathan, ``Remarks on the spatial smoothing step in Co-array MUSIC,'' \textit{IEEE Signal Process. Lett.}, vol. 22, no. 9, pp. 1438-1442, 2015.


\bibitem{KrolikBias}
D.N. Swingler and J. Krolik, ``Source location bias in the coherently focused high-resolution broad-band beamformer,'' \textit{IEEE Trans. Acoust. Speech Signal Process}, vol. 37, no. 1, pp. 143-145, 1989.

\bibitem{noteLRA}
K.C. Indukumar and V.U. Reddy, ``A note on redundancy averaging,'' \textit{IEEE Trans. Signal Process.}, vol. 40, no. 2, pp. 466-469, 1992.

\bibitem{LRAperformance}
M.A. Doron and A.J. Weiss, ``Performance analysis of direction finding using lag redundancy averaging,'' \textit{IEEE Trans. Signal Process.}, vol. 41, no. 3, pp. 1386-1391, 1993.

\bibitem{PillaiACM}
S.U. Pillai and F. Haber, ``Statistical analysis of a high resolution spatial spectrum estimator utilizing an augmented covariance matrix,'' \textit{IEEE Trans. Acoust., Speech Signal Process.,}, vol. 35, no. 11, pp. 1517-1523, 1987.

\bibitem{LiuBuckMDLgap}
Y. Liu and J.R. Buck, ``Wideband Source Enumeration Using Sparse Array Periodogram Averaging in Low Snapshot Scenarios," \textit{J. Acoust. Soc. Am}, In prep. 

\bibitem{kaveh1986statistical}
M. Kaveh and A. Barabell. ``The statistical performance of the MUSIC and the minimum-norm algorithms in resolving plane waves in noise,'' \textit{IEEE Trans. Acoust. Speech Signal Process.}, vol. 34, no. 2, pp. 331-341, 1986. 
\end{thebibliography}
\end{document}